\documentclass[transmag]{IEEEtran}
\usepackage{latexsym}
\usepackage{graphicx}
\usepackage{amsfonts,amssymb,amsmath}
\usepackage{hyperref}
\usepackage{xcolor} 
\usepackage{graphicx}   
\usepackage{amsmath}  
\usepackage{amssymb} 
\usepackage{multirow}
\usepackage{cite}   
\usepackage{breqn}  
\usepackage{algcompatible}
\usepackage{algorithm} 

\newcommand{\qed}{\nobreak \ifvmode \relax \else
      \ifdim\lastskip<1.5em \hskip-\lastskip
      \hskip1.5em plus0em minus0.5em \fi \nobreak
      \vrule height0.75em width0.5em depth0.25em\fi}
      \usepackage[caption=false,font=footnotesize]{subfig}
\graphicspath{{./figures/}} 
\begin{document}
\title{Distributed Massive MIMO for LEO Satellite Networks}
\author{Mohammed~Y.~Abdelsadek,~\IEEEmembership{Senior Member,~IEEE,} Gunes~Karabulut~Kurt,~\IEEEmembership{Senior Member,~IEEE,}
        \\and~Halim~Yanikomeroglu,~\IEEEmembership{Fellow,~IEEE}
\thanks{This article was presented in part at the IEEE International Conference on Communications (ICC) Workshops 2021 in \cite{abdelsadek2021ultra}.\protect\\
 M.~Y.~Abdelsadek and  H.~Yanikomeroglu are with the Department
of Systems and Computer Engineering, Carleton University, Ottawa,
ON K1S 5B6, Canada. Emails: \{mohammedabdelsadek,~halim\}@sce.carleton.ca. M.~Y.~Abdelsadek is also with (on leave) the Department of Electrical Engineering, Assiut University, Assiut 71516, Egypt.}
\thanks{G. Karabulut Kurt is with the Department of Electrical Engineering, Polytechnique Montreal, Montreal, QC  H3T 1J4, Canada. Email: gunes.kurt@polymtl.ca. G. Karabulut Kurt is also an Adjunct Research  Professor in the Department of Systems and Computer Engineering,  Carleton University, Ottawa, ON K1S 5B6, Canada.}}

\IEEEtitleabstractindextext{\begin{abstract}
The ultra-dense deployment of interconnected satellites will characterize future low Earth orbit (LEO) mega-constellations. Exploiting this towards a more efficient satellite network (SatNet), this paper proposes a novel LEO SatNet architecture based on distributed massive multiple-input multiple-output (DM-MIMO) technology allowing ground user terminals to be connected to a cluster of satellites. To this end, we investigate various aspects of DM-MIMO-based satellite network design, the benefits of using this architecture, the associated challenges, and the potential solutions. In addition, we propose a distributed joint power allocation and handover management (D-JPAHM) technique that jointly optimizes the power allocation and handover management processes in a cross-layer manner. This framework aims to maximize the network throughput and minimize the handover rate while considering the quality-of-service (QoS) demands of user terminals and the power capabilities of the satellites.  Moreover, we devise an artificial intelligence (AI)-based solution to efficiently implement the proposed D-JPAHM framework in a manner suitable for real-time operation and the dynamic SatNet environment. To the best of our knowledge, this is the first work to introduce and study DM-MIMO technology in LEO SatNets. Extensive simulation results reveal the superiority of the proposed architecture and solutions compared to conventional approaches in the literature.
\end{abstract}
}
%
%
%

\maketitle
%
\section{Introduction}
\label{sec:introduction}
Satellites are envisaged to play a critical role in future communication networks. Although satellite networks (SatNets) are considered to be one of the most promising solutions for connecting the unconnected in remote and rural areas, they can provide a plethora of additional applications and services elsewhere on earth and in space. In this regard, SatNets can be used to reinforce connectivity solutions in underserved areas \cite{3GPPTR22-822}, and there is a growing interest in utilizing satellite systems to realize ubiquitous Internet of Things (IoT) \cite{centenaro2021survey}. Moreover, SatNets can enable more efficient backhaul services  \cite{shaat2018integrated}, data offloading applications \cite{di2018ultra}, space exploration \cite{gu2020publish}, among other benefits and use cases.

The many use cases of SatNets have drawn attention from network operators, standardization bodies, and manufacturing companies. In this respect, the Third Generation Partnership Project (3GPP)  has been studying the integration of a satellite component in terrestrial fifth-generation (5G) networks through several study items (SIs) in the recent releases.  The SI in \cite{3GPPTR38-811} was implemented in Release 15 to investigate the support of non-terrestrial networks (NTN) (satellites and high-altitude platform stations (HAPS) \cite{kurt2021vision}) in the 5G New Radio (NR). This study has been expanded to several SIs in Releases 16 and 17 to study the following: 1) use cases and satellite access \cite{3GPPTR22-822};   2) integration scenarios and architectural aspects \cite{3GPPTR23-737}; 3) management and monitoring of satellite components \cite{3GPPTR28-808}; and 4) radio access network architecture and interface protocols \cite{3GPPTR38-821}. The focus of standardization efforts in Release~17 is on transparent satellite architecture to enable broadband and backhauling use cases. Future releases will consider the regenerative payload architectures,  IoT, and dual connectivity use cases.

Although geostationary Earth orbit (GEO) satellites have been used for a long time to offer connectivity and broadcast services, there are several issues associated with their services, such as long delay, high path loss, and over-subscription. These drawbacks are in addition to the high costs of manufacturing and deploying GEO satellites. Alternatively, low Earth orbit (LEO) satellites are characterized by low latency and low path loss communications due to lower deployment altitudes (i.e., as low as 300 km compared to $36,000$ km for GEOs). Besides, the deployment costs  of LEOs are much lower than GEOs. This enables LEO SatNets to provide low-cost services with high quality. Therefore, thousands of LEO  satellites are being launched to build mega-constellations orbiting the Earth by 2030, such as those built by SpaceX, Telesat, OneWeb, and Amazon, to name a few.

However, LEO SatNets suffer from several drawbacks due to their motion to  ground user terminals (UTs). This relative motion causes UTs to switch links among different LEO satellites to maintain a network connection. This handover process is implemented at the link-layer and network-layer. The former used for switching over communication links from one satellite to another one in the UT’s visibility. By contrast, a network-layer handover is used for switching higher-layer protocols (e.g., transmission control protocol (TCP), user datagram protocol (UDP)) to a new Internet protocol (IP) address of a UT when it is connected to a different home network due to a satellite handover. The rate of this satellite handover process is high due to the fact that the LEO satellites are only visible to UTs for a few minutes at a time as they pass in orbit. This high rate of satellite handover entails high signalling overhead, throughput losses,  processing delay, data forwarding, and location update issues \cite{darwish2021location}. Therefore, the mobility of LEO satellites can have a significant impact on the network performance, spectrum utilization, and users' quality of service (QoS).  

\subsection{Related Work}
\label{ssec:relatedWork}
Several approaches have been proposed to address challenges associated with the high mobility of  LEO satellites. In \cite{yang2016seamless}, the authors proposed a software-defined networking (SDN) architecture to control LEO satellites using a controller on the ground that connects to the LEOs via a GEO satellite. Along similar lines, the authors in \cite{li2020forecast} proposed an extensible architecture utilizing several layers of terrestrial relays (TRs), HAPSs, LEOs, and GEOs for relay purposes, and they studied the handover procedure among these different systems. However, using GEOs in the network, as investigated in these studies, entails a long-delay segment in the communication cycle that directly impacts the QoS of users and requires coordination between different satellite operators. In addition, these two studies, and other current approaches (e.g., \cite{wu2016graph,wu2019satellite}), focused on the connectivity of UTs with a single satellite in their visibility. Therefore, the UT's service time (i.e., connection time without handover interruption) is limited by the visibility of a single LEO satellite, which is only a few minutes (about 10 minutes in the Iridium system \cite{li2020forecast}, for example). The handover rate is inevitably high, regardless of the adopted handover management technique. Moreover, most of these works are based on old LEO constellations (e.g., Iridium) and do not exploit the features of the new LEO mega-constellations, such as SpaceX's Starlink and Telesat's Lightspeed.

To overcome the challenges of  single-satellite connectivity, the use of multiple-input multiple-output (MIMO) techniques in satellite communications, and in LEO SatNets in particular, has been investigated in a handful of works. Former studies, such as  \cite{schwarz2008optimum,yamashita2005broadband,oh2006analysis},  investigated the connection of ground terminals to two GEO satellites, or to two antennas deployed on a single GEO satellite, for diversity purposes to address fading issues (e.g., in rainy and foggy weathers). As for LEO satellites, the authors in \cite{erdogan2020site} focused on feeder links by studying the connection to multiple ground stations via optical links to realize site diversity in future LEO SatNets. In \cite{feng2020satellite}, the authors adopted a bipartite graph model for the ground gateway stations and the multiple LEO satellites visible to them. They solved this multi-connectivity problem by using maximum matching techniques. However, they utilized basic MIMO techniques (i.e., considering the general case that each ground station node can be connected to multiple satellite nodes and vice versa) and assumed that the ground stations had accurate information about the motion of the LEO satellites in their visibility. In \cite{goto2018leo}, the authors considered a similar MIMO model and analyzed the capacity of the LEO-MIMO links by taking into consideration the Doppler shift due to the motion of the satellites. Nevertheless, these studies considered classical MIMO models to describe the connectivity of ground UTs with multiple satellites without investigating the details of the network architecture, channel estimation, precoding, and interference between users. 

On another front, massive MIMO in LEO SatNets was studied in \cite{you2020massive,you2020leo}. In these works, the authors assumed that the LEO satellites use arrays of uniform planar antennas that can realize massive MIMO. However, due to the line-of-sight (LoS) connection to the ground terminals, the collocated satellite massive MIMO system cannot achieve the desired benefits  as in terrestrial networks if the ground users are not sufficiently separated due to the  so-called ``unfavourable propagation" environment \cite{arnau2014dissection} that results in a ``keyhole channel" matrix \cite{schwarz2019mimo}. Therefore, collocated massive MIMO would not be suitable for single-user MIMO scenarios (sending multiple layers to the same user), for instance.

One significant difference between future LEO SatNets and old LEO constellations is ultra-dense deployment. That is, future LEO mega-constellations will include thousands of satellites. For example, SpaceX is planning to deploy $30,000$ LEO satellites for their second-generation constellation in addition to the current plan of around $12,000$ satellites \cite{FCC2021spacex}. This means that multiple LEO satellites will be visible to ground UTs simultaneously, which will open the door for advanced distributed MIMO techniques, such as cell-free massive MIMO (CF-mMIMO). 

CF-mMIMO was recently proposed for next-generation terrestrial networks that build on coordinated multi-point (CoMP) and network MIMO techniques for large spectral efficiency, power efficiency, and network flexibility gains \cite{ngo2017cell}. In terrestrial CF-mMIMO, multiple access points can be used to cooperatively communicate with users in a cell-free manner.  This technique can be utilized in future LEO satellite networks exploiting an ultra-dense deployment, very high-speed inter-satellite links (ISLs), and LoS connections with ground UTs. Moreover, the CF-mMIMO architecture enables the cross-layer design to jointly optimize the upper and lower layers of networking. This  cross-layer design is of utmost importance to  LEO SatNets given that the network nodes are moving (i.e., the network topology is dynamic) and all links (with UTs, gateways, and other satellites) are  wireless (radio frequency (RF) or free-space optical (FSO)). This means that the design of the lower layers has a significant impact on the performance of the upper ones. Therefore, CF-mMIMO can be leveraged for an efficient, resilient satellite network.

Noting that CF-mMIMO is a terrestrial technology that belongs to the distributed MIMO techniques, in this paper, we focus on the generalization of CF-mMIMO and propose a distributed massive MIMO (DM-MIMO) LEO satellite network architecture. Besides, we investigate the cross-layer design and an artificial intelligence (AI)-based implementation. To the best of our knowledge, this is the first work to introduce and study a DM-MIMO approach in LEO constellations.

%
\subsection{Paper Contributions and Structure}
\label{ssec:contributions}
The major contributions of this paper can be summarized as follows:
\begin{itemize}
\item We propose a LEO SatNet architecture based on DM-MIMO techniques. More specifically, we investigate the network topology, required ISLs, duplexing mode, beamforming, power control, frequency reuse, and handover management strategies. Moreover, we highlight the benefits of using this architecture and investigate the associated challenges and potential solutions to realize it.
\item We develop an optimized cross-layer design framework based on the proposed DM-MIMO-based architecture, such that the power allocation and handover management processes are jointly optimized. For this purpose, we describe the channel model, uplink training and channel estimation, downlink data transmission, and formulate a novel multi-objective optimization problem. It should be noted that it is not straightforward to establish a combined channel and data transmission models based on those from cell-free massive MIMO and satellite communications. In the optimization problem, the aggregate throughput is maximized while minimizing the handover rate. We refer to this optimized cross-layer design as the distributed joint power allocation and handover management (D-JPAHM) technique. 
\item We introduce an AI-based implementation for the developed cross-layer control framework, which can be used in practical satellite systems. For this purpose, we leverage deep learning to provide accurate predictions for the solution of the formulated multi-objective optimization problem without actually solving the problem.  Deep learning provides several benefits compared to traditional optimization-based and heuristic approaches. For instance, it exploits offline computations to reduce the computational complexity of the online operation. It also adapts to changing environments, which is vital for the dynamic satellite network. Furthermore, the scalability and support of distributed data processing and storage provided by deep learning techniques are crucial for LEO satellites that are less capable compared to GEO satellites.
\item We conduct extensive simulations to evaluate the performance of the proposed architecture, cross-layer design, and AI-based implementation. In addition, we compare the performance with that of conventional approaches and architectures from the literature. The simulation results show the superiority of the proposed DM-MIMO-based architecture and solutions compared to the traditional single-connectivity approach.
\end{itemize}
%

The remainder of this paper is organized as follows. In Section \ref{sec:architecture}, the proposed DM-MIMO-based LEO SatNet architecture is discussed, and several network design aspects are investigated. Then, in Section \ref{sec:BenefitsAndChallenges}, we highlight the benefits of using this architecture,  the associated challenges and potential solutions. In Section  \ref{sec:crosslayerDesign}, the optimized cross-layer design is detailed. 
In Section \ref{sec:DLApproach}, we discuss the AI-driven approach for implementing the proposed cross-layer optimization framework. In Section \ref{sec:Results}, we present and discuss the results of the simulations to evaluate the performance of the proposed DM-MIMO-based architecture and solutions in comparison with that of traditional single satellite connectivity. Finally, we conclude the paper in Section \ref{sec:Conclusions}.

%
\section{DM-MIMO-Based SatNets}
\label{sec:architecture}
%
\subsection{Architecture}
\begin{figure}
\centering    
\includegraphics[width=90mm]{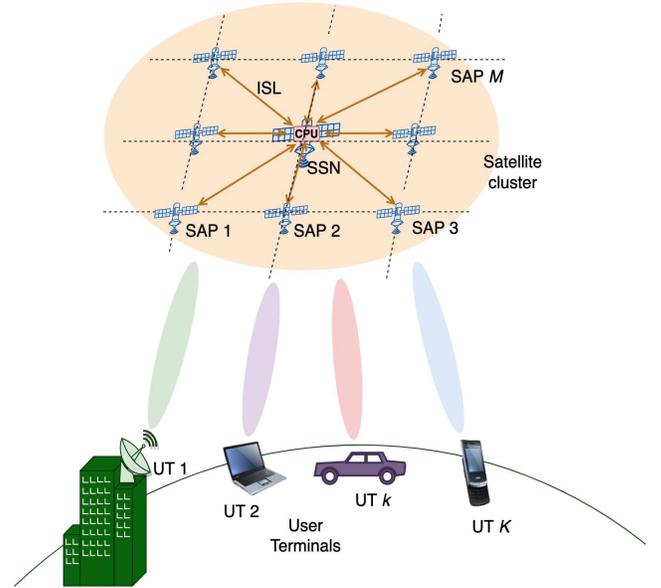}
\caption{DM-MIMO-based LEO SatNet. The user terminals are connected to a cluster of LEO satellite access points (SAPs) that are controlled by a central processing unit (CPU), which is deployed on a super satellite node (SSN).}
\label{fig:architecture}
\end{figure}
\begin{figure}
\centering   
\includegraphics[width=80mm]{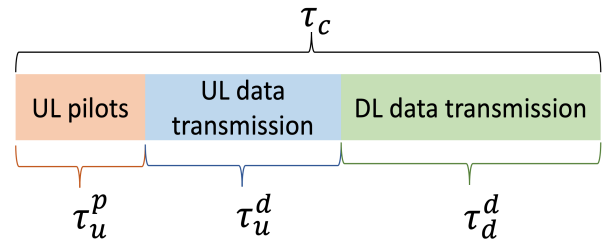}
\caption{The TDD frame is composed of three parts used for uplink pilot transmission, uplink data transmission, and downlink data transmission.}
\label{fig:TDDFrame}
\end{figure}

Fig.~\ref{fig:architecture} shows the proposed DM-MIMO-based LEO SatNet architecture. As we can see, the satellites are divided into clusters. To be consistent with the terrestrial CF-mMIMO terminology, each satellite in the cluster is called a satellite access point (SAP). In the event that a satellite has multiple antennas, each antenna would be considered a separate SAP. These SAPs are connected to a central processing unit (CPU) through ISLs for fronthaul purposes. This CPU can be deployed on a central satellite with more advanced computing capabilities. These central satellites are called super satellite nodes (SSNs). 

%
\subsection{TDD Operation}
To exploit the reciprocity of the uplink (UL) and downlink (DL) channels, time division duplexing (TDD) is the adopted duplexing mode in CF-mMIMO systems \cite{ngo2017cell}. For the purpose of channel estimation, both UL and DL pilots can be used \cite{interdonato2019ubiquitous}. However, in most CF-mMIMO studies (e.g., \cite{ngo2017cell,chen2018channel,bjornson2019making}), only UL pilots are considered. This is also suitable for SatNets because the users do not need to estimate their effective channel gain and to use most of the TDD frame for data transmission. It is important to mention that the propagation delay and Doppler shift can be compensated in the time and frequency synchronization processes, given that the satellites move in a predetermined orbit, and their motion information can be known a priori \cite{peters2020doppler}. Therefore, TDD was adopted in several satellite LEO communication studies \cite{xia2020maritime} and systems \cite{monte1991mobile,tan2019positioning}. Accordingly, the TDD frame can be structured, as shown in Fig.~\ref{fig:TDDFrame}.

The channel coherence interval is defined as the time-frequency interval during which the channel characteristics can be considered static. This coherence interval depends on the channel condition, the mobility of the satellite and the UT, and the carrier frequency. The coherence interval samples--or channel uses--designated $\tau_c$, are divided into three parts: the initial $\tau^p_u$ samples are used for UL pilot transmission, the next $\tau^d_u$ samples are used for UL data transmission, and the last $\tau^d_d$ samples are reserved for DL data transmission. It is worth mentioning that the guard intervals are excluded from this coherence time interval. Utilizing the UL pilots, all the UL channels are estimated at the SAPs locally without forwarding them to the CPU. This supports the scalability of the network, since the signalling overhead is independent of the number of SAPs. Due to reciprocity, these channel estimates are valid for the DL direction as well. Therefore, the estimated channels are used for DL data precoding and UL data detection.

%
\subsection{Radio Resource Allocation}
Efficient radio resource allocation techniques can exploit the advantages of the proposed architecture (including transmit and receive diversity) to  achieve higher throughput for the connected terminals, ensuring their QoS satisfaction, minimizing interference, and minimizing the handover rate. In this subsection, we discuss different aspects of resource allocation for the proposed DM-MIMO-based SatNets architecture.

\subsubsection{Pilot assignment}
UTs can be assigned mutually orthogonal UL pilots to minimize the interference between them. However, this requires the number of UL training samples, $\tau^p_u$, to be more than the number of connected UTs, which is difficult in SatNets due to the large number of connected UTs. Therefore, every subset of the UTs can be assigned one pilot from the mutually orthogonal pilot set. This results in what is known as pilot contamination, which needs to be taken into consideration while designing the resource allocation procedure. The pilot assignment can be implemented locally at the SAPs in a distributed manner or centrally at the CPU. The pilot assignment information can be transmitted to the UTs over the random access channel during the random access process.
\subsubsection{Beamforming}
Several beamforming techniques can be utilized for this DM-MIMO-based satellite networks. One of the widely used schemes in the literature (e.g., in \cite{ngo2017cell,chen2018channel,ozdogan2019performance}), is the maximum ratio processing (i.e., conjugate beamforming in the DL direction and matched filtering in the UL). This method can be employed to exploit the distributed channel estimation at the SAPs, which is considered one of the major benefits of using DM-MIMO, as this reduces the computational complexity and the required fronthaul signalling between the SAPs and the CPU \cite{ngo2017cell}. However, other centralized beamforming techniques, such as zero-forcing (ZF) and minimum mean square error (MMSE) can be used \cite{nayebi2017precoding,bjornson2019making}. These schemes can be utilized to optimize the beamforming design at a global manner to improve performance. However, this centralized operation requires more fronthaul signalling between the SAPs and CPU to allow the CPU to collect the channel parameters from the SAPs.

\subsubsection{Frequency reuse}
By using precoding in the DL and maximum ratio combining for the UL transmissions, full frequency reuse (FFR) can be realized instead of using conventional four colour frequency reuse patterns to mitigate the interference between the spot beams \cite{vazquez2016precoding}. This leads to efficient spectrum usage since the whole allocated spectrum can be used anytime and anywhere while minimizing the interference among users. 

\subsubsection{Power control}
Power control plays an essential role in optimizing the cooperative transmission and reception of SAPs to maximize the network throughput and ensure users satisfaction. The power allocation should consider the interference between the UTs, the pilot assignment, and the achievable data rates. In addition, power allocation can be optimized to maximize the service time of the UTs, thereby minimizing the handover rate. This is the focus of Section \ref{sec:crosslayerDesign}.
%
\subsection{Handover Management}
\label{ssec:handoverManagement}
\begin{figure*}[t!]
\centering     
\subfloat[]{\label{fig:TraditionalHO}\includegraphics[height=7cm, width=3.2in]{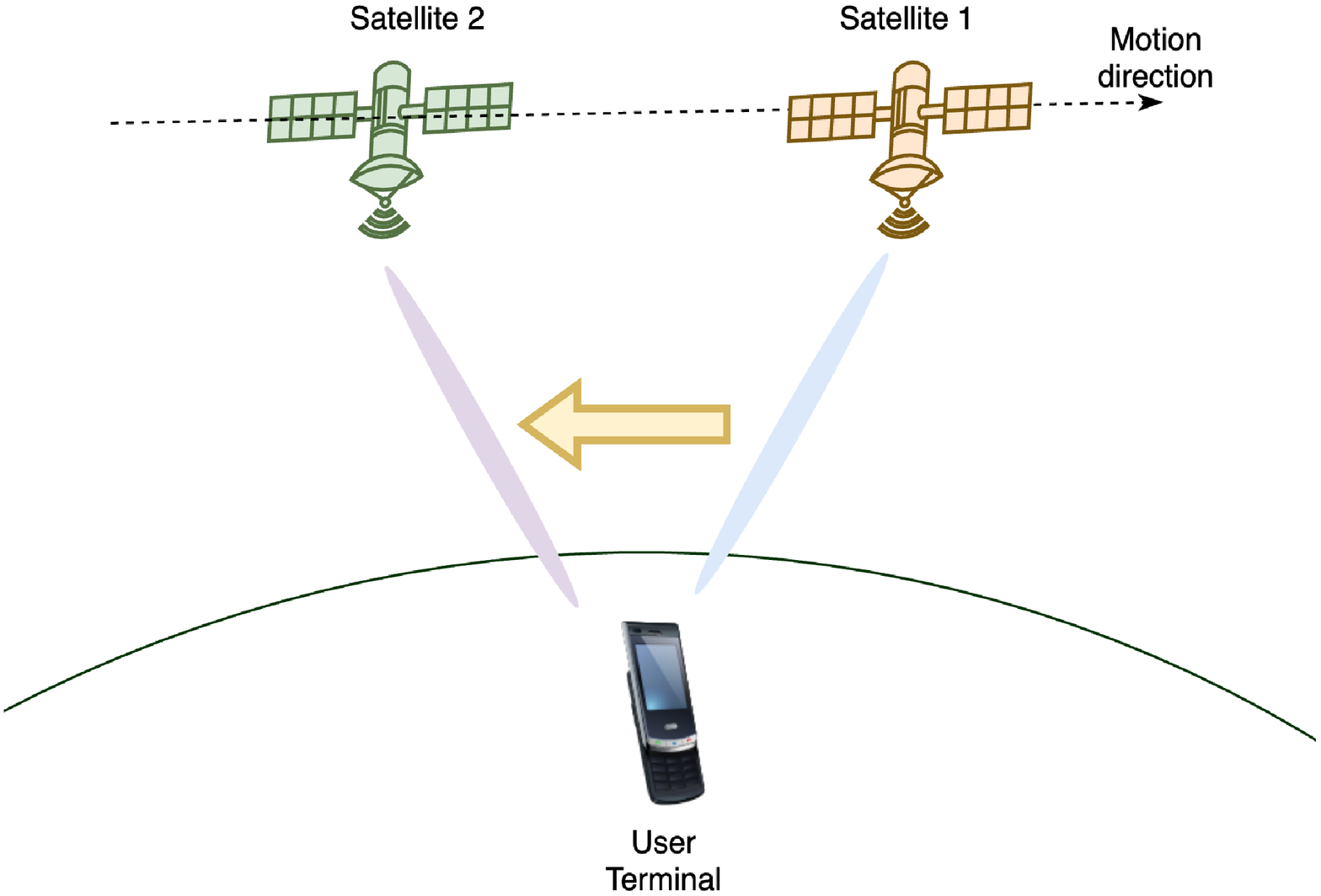}}
\hfil
\subfloat[]{\label{fig:ClusterHO}\includegraphics[height=7cm,width=3.2in]{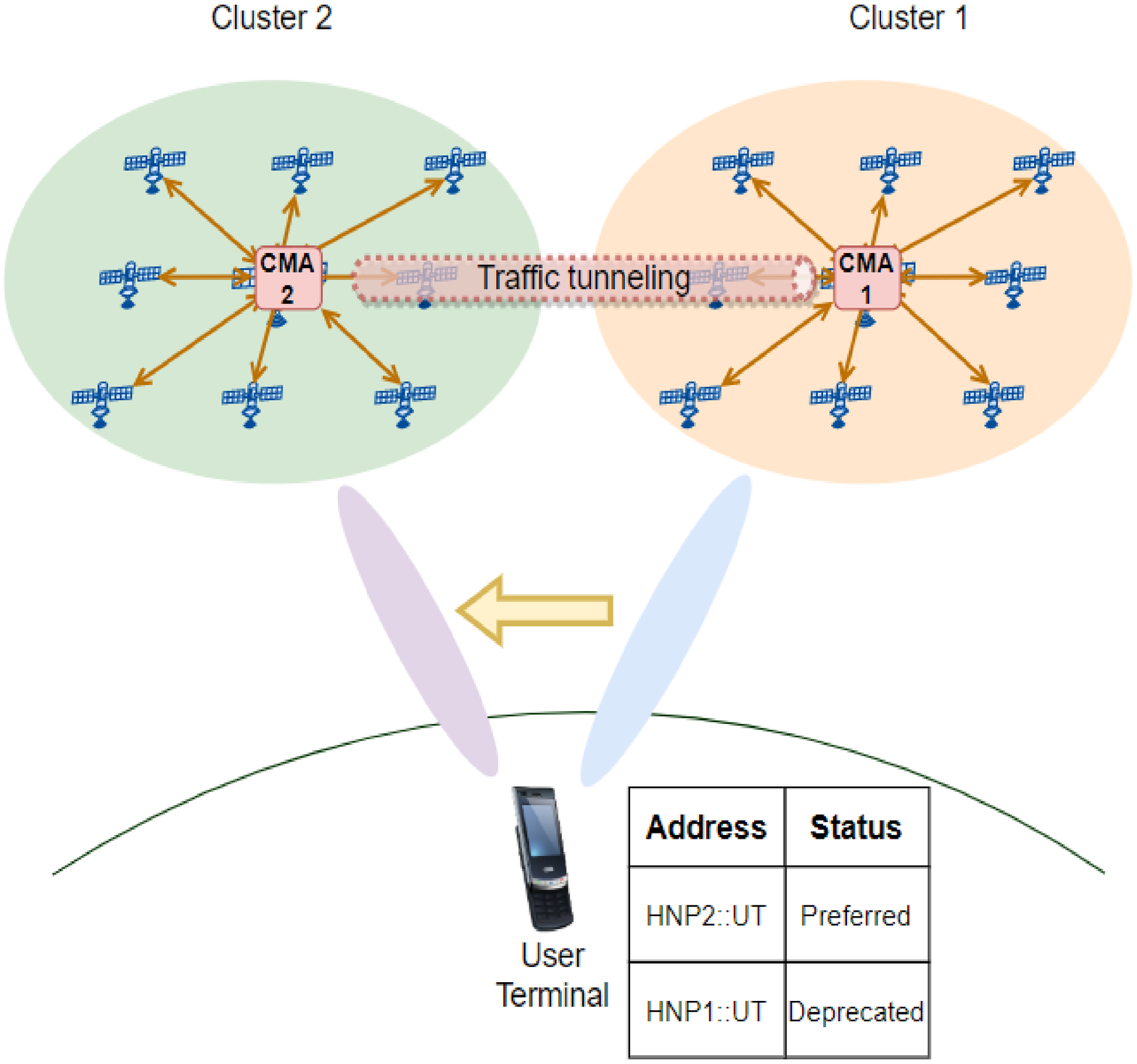}}
\caption{Single-satellite versus cluster handover. In (a), the user terminals switch their connections among individual satellites based on their visibility. In (b), the handover is processed at a cluster level which reduces the handover rate. }
\label{fig:SatVsClusterHO}
\end{figure*}

The traditional satellite handover process (i.e., based on single satellite connectivity) is depicted in Fig.~\ref{fig:TraditionalHO}. In this case, when the signal level is below a certain threshold, the link is switched to the next LEO satellite in the cone visibility of the UT. This can be accomplished by using the satellite reference signals that are broadcast by the satellite. In the network layer, handover is required to forward the data arriving for the old address to the new home network, since the UT is given a new IP address in the new home network. As discussed above, the  service time in this case is limited by the satellite visibility, which is a few minutes in LEO SatNets. However, using the proposed DM-MIMO architecture, the UT is connected to a cluster of satellites or SAPs. Consequently, the service time is limited by the visibility of the target cluster, which is longer than that of a single satellite. In addition, the resource allocation process can be optimized such that the service time is maximized. This minimizes the handover rate, associated losses, processing delays, and signalling overhead.  In what follows, we discuss the link-layer and network-layer handover processes based on the proposed DM-MIMO architecture.
%
 
\subsubsection{Link-layer handover}
In DM-MIMO-based SatNets, power allocation can be adjusted such that the service times of the ground UTs are maximized. This is because the UTs are served by all SAPs in the serving cluster. Therefore, the cooperative transmission of those SAPs can compensate for the signal level decaying due to the movement of the satellites. In addition, in the UL direction, the data is decoded on the basis of the received signals by all SAPs in the cluster. Nevertheless, a cluster handover is required to switch to the next cluster when resource allocation cannot satisfy the UT's minimum required data rate level. This can be detected while allocating the radio resources (i.e., the transmit power), as detailed in Section \ref{sec:crosslayerDesign}. Besides, since the next serving cluster is known, the handover decision can be confirmed by the next cluster that detects the UL pilot from the  ground UT by its edge SAPs. Fig.~\ref{fig:ClusterHO} shows the handover between satellite clusters when the minimum QoS level cannot be guaranteed by the old cluster's established link due to moving far from the UT.

\subsubsection{Network-layer handover}
When link handover is triggered and implemented, a network-layer handover is required to assign a new address to the UT and forward the incoming data to the new home network. In terrestrial IP networks, several mobility management protocols are adopted, such as Mobile IPv6 (MIPv6) \cite{IETFRFC6275} and Proxy Mobile IPv6 (PMIPv6) \cite{IETFRFC5213}. In these protocols, mobility anchors are used to establish tunnels to forward the data to the new network and update the binding cash. In the proposed DM-MIMO architecture, the addressing issues can be tackled by using cluster mobility anchors (CMAs) that are located in the SSNs of the clusters (i.e., along with the CPUs). When the link handover is implemented, the CMA of the new cluster provides a new home network prefix (HNP) to the UT and establishes a bidirectional tunnel with the old serving CMA such that the data are forwarded to the new cluster, as shown in Fig.~\ref{fig:ClusterHO}. In this scheme, the old IP address of the UT (\texttt{HNP1::UT}) becomes Deprecated, and the newly assigned address (\texttt{HNP2::UT})  becomes Preferred.

\begin{figure}
\centering   
\includegraphics[width=90mm]{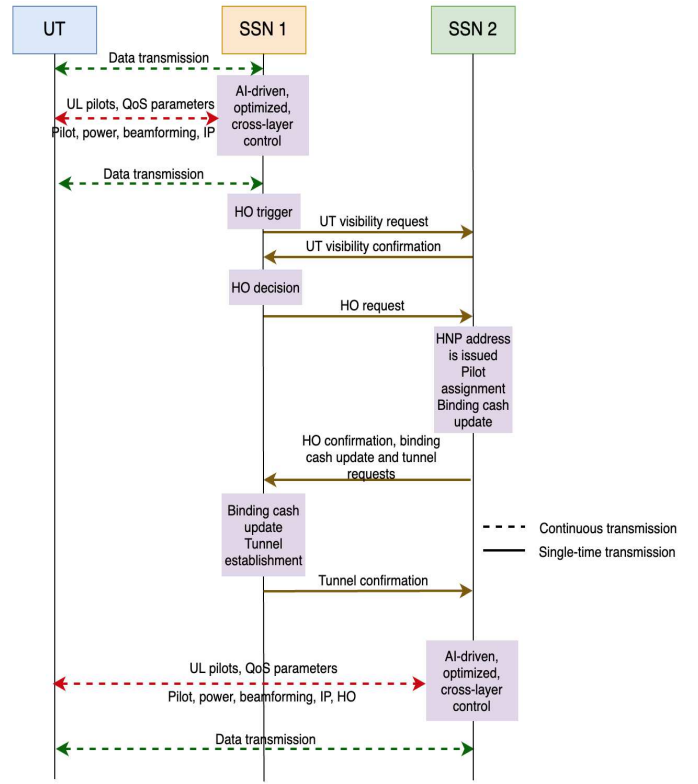}
\caption{Handover procedure. When a handover is triggered and the visibility of the next cluster is confirmed, a network-layer handover procedure is implemented to forward the packets to the new IP address in the new cluster.}
\label{fig:HOProcedure}
\end{figure}

The link and network handover procedure for the proposed DM-MIMO can be implemented as depicted in Fig.~\ref{fig:HOProcedure}. When a handover is triggered at the link level and the visibility of the UT by the next cluster is confirmed, a network handover procedure is implemented to create a tunnel between the old and new clusters. This allows the packets arriving for the UT with the old IP address to be directed to the new cluster. Then, cross-layer control is implemented to assign new pilot, power, beamforming vector, ... etc., to the UT in the new cluster to resume the data transmission via that cluster of SAPs. The complete signal diagram is shown in Fig.~\ref{fig:HOProcedure}.
%

%
%
\section{Benefits, Challenges, and Potential Solutions}
\label{sec:BenefitsAndChallenges}
Based on the previous discussion of the different design aspects of DM-MIMO-based SatNets, we highlight the major benefits of using this architecture, the main challenges associated, and potential solutions to address these challenges in this section.

\subsection{Benefits}
The major benefits of using DM-MIMO in LEO SatNets can be summarized as follows:
\begin{itemize}
\item \textit{Improved throughput:} As indicated earlier, the dominant LoS communication between the ground UTs and LEO satellites imposes limitations on the use of MIMO techniques in LEO SatNets. This is due to the unfavourable propagation environment that lowers the rank of the channel matrix between the UTs and satellites  \cite{arnau2014dissection}. Therefore, this limits the MIMO gain to the keyhole capacity as investigated in \cite{schwarz2019mimo}, except if proper user grouping and spacing are employed to alleviate this effect. The use of DM-MIMO realizes this spatial distribution among the serving satellites and could address the collocation issues. This is in addition to the full frequency reuse that enhances spectral efficiency. Therefore, the use of DM-MIMO could significantly improve the satellite network throughput.

\item \textit{Lower handover rate:} As discussed in Section \ref{ssec:handoverManagement}, the use of DM-MIMO-based architecture enables the UTs to be served by a cluster of LEO satellites. Therefore, this extends the UT's service time to the visibility period of a cluster of satellites instead of a single one. This directly reduces the handover rate of UTs and the issues associated with a high handover rate, such as signalling overhead, processing delay, data forwarding, tunnelling, IP addressing issues, and location update.

\item \textit{Flexible operation:} The concept of CPUs and SAPs is similar to centralized units (CUs) and distributed units (DUs) that are used in the context of open radio access networks (O-RAN), as discussed in \cite{ranjbar2022cell}. This opens the door to the advanced flexible O-RAN techniques  \cite{abdalla2022toward} to be utilized in LEO SatNets.

\item \textit{Cross-layer design:} As indicated in Section \ref{sec:introduction}, the nodes of the satellite network are moving and have wireless links among them and to the users on the ground. Therefore, the performance of lower layers highly impacts the upper ones. The use of DM-MIMO enables the cross-layer design of SatNets to jointly optimize upper and lower layers, which provides several benefits to SatNets \cite{giambene2006cross}. In this regard, a cross-layer framework is proposed in Section \ref{sec:crosslayerDesign} to show the benefits of using this design.
\end{itemize}

\subsection{Challenges and Potential Solutions}
\label{ssec:ArchChallenges}
Although the DM-MIMO-based architecture provides several benefits to LEO SatNets, several challenges need to be addressed to leverage this technology in SatNets. The major challenges and potential solutions can be summarized as follows:
\begin{itemize}
\item \textit{Dynamic topology:} Due to the dynamic environment of LEO satellites (due to movement at different directions and at different altitudes), the topology of network nodes changes with time. This entails issues with defining the cluster that serves the UTs (i.e., the number of SAPs and which SAPs serve each UT). However, this cluster formation process can be optimized by the SSN on the basis of different performance objectives and associated deployment costs. For instance, this depends on the amount of traffic per spot (e.g., hotspots should be served with a larger number of SAPs compared to cold spots), the capabilities of the serving SAPs (e.g., the power and computational resources), and the kind of traffic (e.g., narrowband versus broadband). An example of such a dynamic clustering approach is proposed in \cite{bjornson2020scalable} for terrestrial cell-free massive MIMO networks. This concept can be adapted to DM-MIMO-based SatNets. Moreover, user grouping (see \cite{storek2017fair} for instance) can be employed to assign groups of ground UTs to dynamic clusters of SAPs to optimize user-cluster association.

\item\textit{Synchronization:} To achieve such cooperative transmission between a cluster of satellites, synchronization in time, frequency, and phase is required. Several time synchronization techniques have been proposed in the literature to align the signals at the symbol level and achieve coherent transmission (see \cite{pan2008research,jiuling2018approach} for instance). Besides, frequency and phase synchronization has been widely studied in the literature to achieve cooperative transmission or sensing as discussed in \cite{jayaprakasam2017distributed}. These synchronization techniques can be utilized and adapted to be applied in the considered scenario of DM-MIMO-based LEO SatNets.

\item \textit{Outdated channel information:} As discussed in Section \ref{sec:architecture}, the channel knowledge (which is vital to several network management functionalities) is based on the estimated channel utilizing the UL pilots. However, due to the long propagation delay (around $7.9$ ms for LEOs), this channel information could be outdated. To overcome this issue, intelligent machine learning techniques can be utilized to estimate the channel while taking into account the propagation delay. An example of this approach is studied in \cite{zhang2021deep}, where the authors propose a deep learning model that is based on long short term with memory (LSTM) to predict the channels while exploiting their correlation to overcome channel aging issues in LEO SatNets.

\item \textit{Complexity of SSNs:} The SSNs are used to control the SAPs in their clusters. Therefore, this requires additional computing and power resources for these satellites to be able to perform such control tasks. To overcome this issue, distributed processing can be utilized as much as possible. For example, as discussed in Section \ref{sec:architecture}, several network management functions (e.g., beamforming and pilot assignment) can be implemented in a distributed manner to reduce the computations on the SSNs. In addition, software operation can be exploited to flexibly implement the functions of the CPU without requiring extra hardware at the SSNs.

\item \textit{Signalling overhead:}  To implement the cooperative operation of SAPs, additional signalling between the CPU and SAPs is required. To overcome this signalling overhead, the exchange of the information between the SAPs and the CPU should be minimized; for example, by restricting it to payload data and power control coefficients as discussed in \cite{ngo2017cell}. Besides, the use of high-speed FSO communications \cite{chaudhry2020free}, which enables the information to be exchanged at very high rate with low latency, can be utilized to overcome this  issue.  
\end{itemize}

%
%
\section{Optimized Cross-Layer Design}
\label{sec:crosslayerDesign}
For the cross-layer design of the power allocation and handover management processes in the proposed DM-MIMO-based SatNet architecture, we start by discussing the channel model and estimation.  We then derive the achievable data rates and formulate the optimization problem. The frequently used symbols are summarized in Table \ref{tab:UsedSymbols}.
\begin{table}[t]
\renewcommand{\arraystretch}{1.3}
\caption{Frequently Used Symbols}
\label{tab:UsedSymbols} 
\centering
\begin{tabular}{llc}
\hline 
\textbf{Symbol} & \textbf{Description} \\ 
\hline 
$\mathcal{K},~\mathcal{M}$& Sets of UTs and SAPs, respectively \\
$K,~M$& Cardinalities of $\mathcal{K},~\mathcal{M}$, respectively\\
$\tau_c,~\tau_u^p$& Length of coherence interval and UL pilot, respectively\\
$\tau_u^d,~\tau_d^d$& Length of UL and DL data transmissions, respectively\\
$q_k,~p_{m,k}$ & UL pilot power and DL power factor of SAP $m$ to UT $k$\\
$h_{m,k}$ & Channel coefficient between UT $k$ and SAP $m$\\
$\hat{h}_{m,k}$ & Estimated channel coefficient between UT $k$ and SAP $m$\\
$L_{m,k}$ & Large-scale fading and losses between UT $k$ and SAP $m$\\
$R_k$ & Achievable DL data rate in bps/Hz for UT $k$\\
$R_k^{min}$ & Minimum rate of UT $k$ in bps/Hz\\
$P_m^{max}$ & Maximum power factor of SAP $m$\\
$\alpha$ & Handover rate minimization priority factor \\
\hline
\end{tabular}
\end{table}
%

\subsection{Channel Model}
We consider a cluster of LEO satellites that includes a set of $M$ SAPs indexed by $\mathcal{M}=\{1,~2,\cdots,~m,\cdots,~M\}$. This cluster serves a set of single-antenna ground UTs set, indexed by $\mathcal{K}=\{1,~2,\cdots,~k,\cdots,~K\}$. Assume that the channel conditions are static in a coherence time interval of $\tau_c$ samples. Due to the strong LoS component between the UTs and SAPs, the channel between the $k$th UT and the $m$th SAP is modelled as Rician and can be calculated by \cite{ngo2018performance}
\begin{align}
\label{eq:channelCoeff1}
h_{m,k}=\sqrt{L_{m,k}}\left(\sqrt{\frac{\kappa_{m,k}}{\kappa_{m,k}+1}} h'_{m,k}+\sqrt{\frac{1}{\kappa_{m,k}+1}}h''_{m,k} \right),
\end{align}
where $\kappa_{m,k}$ is the Rician K-factor, $h'_{m,k}$ and $h''_{m,k}$ represent the LoS and non-LoS (NLoS) components, respectively. The large scale fading and losses are represented by $L_{m,k}=\newline  10^{-(L^{dist}_{m,k}+L^{shad}_{m,k}+L^{angl}_{m,k})/10}$, where $L^{dist}_{m,k}$ is the power loss (in dB) due to distance between UT $k$ and SAP $m$, $L^{shad}_{m,k} \sim\mathcal{N}(0,\sigma_{sh}^2)$ is the shadowing attenuation (in dB) with variance $\sigma_{sh}^2$, and $L^{angl}_{m,k}$ is the loss due to the boresight angle and can be calculated (in dB) by \cite{li2020hierarchical}
\begin{align}
L^{angl}_{m,k}=-10 \log_{10} \left( \cos(\theta_{m,k})^{\eta} \frac{32 \log 2}{2\left( 2~\text{arccos}(\sqrt[\eta]{0.5}) \right)^2} \right),
\end{align}
where $\theta_{m,k}$ is the boresight angle between the $k$th UT and the $m$th SAP, and $\eta$ is the antenna factor determining the coverage radius, assuming that the aperture efficiency is unity.

Suppose that the NLoS component, $h''_{m,k}$, is a Rayleigh random variable, i.e., $h''_{m,k}\sim \mathcal{C}\mathcal{N}(0,1)$. The LoS component is given by $h'_{m,k}=e^{j\phi_{m,k}}$, where $\phi_{m,k}\sim \mathcal{U}[-\pi,\pi]$ is a uniform random variable that represents the phase shift due to the mobility of the SAP and UT and propagation delay. 

For simplicity, we rewrite (\ref{eq:channelCoeff1}) as follows:
\begin{align}
\label{eq:channelCoeff2}
h_{m,k}=\sqrt{\beta_{m,k}}e^{j\phi_{m,k}} + \tilde{h}_{m,k},
\end{align}
where
\begin{align}
\label{eq:channelCoeff3}
\beta_{m,k} = \frac{\kappa_{m,k}}{\kappa_{m,k}+1}L_{m,k}.
\end{align}
In (\ref{eq:channelCoeff2}), $\tilde{h}_{m,k}\sim \mathcal{C}\mathcal{N}(0, \lambda_{m,k})$, $\lambda_{m,k}=L_{m,k}/(\kappa_{m,k}+1)$, and $\beta_{m,k}\in \mathbb{R}$. Therefore, $\beta_{m,k}$ and $\lambda_{m,k}$ are changing slowly  \cite{feng2017coordinated} compared to small-scale fading that changes instantaneously. Since they mainly depend on the UT's position, they can be calculated a priori. In addition, we assume that the propagation delay and Doppler shift are compensated in the time and frequency synchronization processes. This is a reasonable assumption since they primarily depend on the satellite and user location and velocity, which can be determined given the satellite orbital information \cite{peters2020doppler}. In addition, other advanced techniques can be utilized to compensate for Doppler offset in real-time  and achieve symbol synchronization without a priori satellite information (e.g., \cite{peters2020doppler,li2016fast}).
%
\subsection{Uplink Training and Channel Estimation}
As discussed in Section \ref{sec:architecture}, TDD is used with the frame structure shown in Fig.~\ref{fig:TDDFrame}. Therefore, to estimate the UL channels at the SAPs, every UT transmits a pilot on the initial $\tau_u^p$ samples of the coherence block. Since we assume that the number of UTs is larger than the number of mutual orthogonal pilots (i.e., $K>\tau_u^p$), every subset of UTs is assigned the same pilot. The subset of UTs that are assigned the same pilot as UT $k$ is denoted by $\mathcal{C}_k$. Define $\sqrt{q_k}\psi_k\in \mathbb{C}^{\tau_u^p\times 1}$ as the $\tau_u^p$-length pilot sequence transmitted by the $k$th UT, where $q_k$ is the pilot power and $\psi_k^H\psi_k=||\psi_k||^2=\tau_u^p$. Therefore, the received signal vector at the $m$th SAP, $\textbf{y}_m^p\in \mathbb{C}^{\tau_u^p\times 1}$, from all $K$ UTs' pilot transmissions is given thus:
\begin{align}
\textbf{y}_m^p=\sum_{k=1}^K \sqrt{q_k}h_{m,k}\psi_k + \textbf{n}_m^p,
\end{align}
where $\textbf{n}_m^p\sim \mathcal{CN}(\textbf{0}_{\tau_u^p},\sigma_{n^p}^2\textbf{I}_{\tau_u^p})$ is the additive white Gaussian noise (AWGN) vector.

To estimate the UL channel of UT $k$, sufficient statistics are derived from the received signal by calculating the inner product between the received signal vector, $\textbf{y}_m^p$, and $\psi_k$, as follows:
\begin{align}
y_{m,k}^p&=\psi_k^H \textbf{y}_m^p=\sum_{k'=1}^K \sqrt{q_{k'}}h_{m,k'}\psi_k^H \psi_{k'}+\psi_k^H \textbf{n}_m^p\\
&= \sqrt{q_k}\tau_u^ph_{m,k}+\sum_{k'\in\mathcal{C}_k\backslash \{k\}} \sqrt{q_{k'}}h_{m,k'}\tau_u^p+\psi_k^H \textbf{n}_m^p.
\end{align}
This is because
\begin{align}
\psi_k^H \psi_{k'}= \begin{cases}
\tau_u^p, & k'\in \mathcal{C}_k\\
0, & \text{otherwise}\\
\end{cases}.
\end{align}

This statistic can be used to estimate the UL channel, $h_{m,k}$, at the $m$th SAP using techniques such as MMSE and linear MMSE (LMMSE) estimators. We assume that a phase-aware MMSE channel estimator is used. Therefore, the estimated UL channel can be given accordingly \cite{ozdogan2019performance}:
\begin{align}
\label{eq:estimatedChannel}
\hat{h}_{m,k}&=\sqrt{\beta_{m,k}}e^{j\phi_{m,k}}+\frac{\sqrt{q_k}\lambda_{m,k}(y^p_{m,k}-\bar{y}^p_{m,k})}{\gamma_{m,k}},\\
\bar{y}^p_{m,k}&=\sum_{k'\in\mathcal{C}_k} \sqrt{q_{k'}} \tau_u^p \sqrt{\beta_{m,k'}}e^{j\phi_{m,k'}},\\
\gamma_{m,k}&= \sum_{k'\in \mathcal{C}_k}q_{k'}\tau_u^p \lambda_{m,k'}+\sigma_{n^p}^2,
\end{align}
with the following statistics
\begin{align}
\mathbb{E}\{\hat{h}_{m,k}|\phi_{m,k}\}&=\sqrt{\beta_{m,k}}e^{j\phi_{m,k}},\\
\text{Var}\{\hat{h}_{m,k}|\phi_{m,k}\}&= \frac{q_k\tau_u^{p}\lambda_{m,k}^2}{\gamma_{m,k}},
\end{align}
where $\mathbb{E}\{\cdot\},~\text{and} ~\text{Var}\{\cdot\}$ are the expectation and variance operators, respectively.
%
\subsection{Downlink Data Transmission}
Given that most of the traffic is in the DL direction, we consider the DL power allocation and manage the handover process. In the DL, the SAPs transmit the same symbol to the UT in a cooperative manner. Assume that the symbol to be sent to UT $k$ is $s_k\in \mathbb{C}$. Every symbol is precoded by a precoding vector $\textbf{v}_k=[v_{1,k},v_{2,k},\cdots,v_{M,k}]^T$, where $v_{m,k}\in \mathbb{C}$. Therefore, if the signal vector to be sent to the $K$ UTs is $\textbf{s}=[s_1,s_2,\cdots,s_K]^T$, then the signal vector to be transmitted by the $M$ SAPs is given thus:
\begin{align}
\textbf{x}&=\textbf{V}\textbf{s}= \textbf{v}_1s_1+\textbf{v}_2s_2+\cdots+\textbf{v}_Ks_K,
\end{align} 
where $\textbf{V}=[\textbf{v}_1,\textbf{v}_2,\cdots,\textbf{v}_K]$ is an $M\times K$ matrix. 

Therefore, the signal received by the $k$th UT can be calculated by
\begin{align}
y_k&=\textbf{h}_k^H\textbf{x}\\
&=\textbf{h}_k^H\textbf{v}_ks_k+\sum_{k'\in \mathcal{K}\backslash k}\textbf{h}_k^H\textbf{v}_{k'}s_{k'}+n_k,
\end{align}
where $\textbf{h}_k=[h_{1,k},h_{2,k},\cdots,h_{M,k}]^T$ and $n_k\sim\mathcal{CN}(0,\sigma^2_n)$ is the AWGN noise. Assuming that the UT approximates the precoded channel by the average value $\mathbb{E}\{\textbf{v}_k^H\textbf{h}_k\}$, the signal-to-interference and noise (SINR) can be calculated by \cite{ozdogan2019performance}
\begin{align}
\label{eq:SINR}
\text{SINR}_k=\frac{|\mathbb{E}\{\textbf{v}_k^H\textbf{h}_k\}|^2}{\sum_{i=1}^K\mathbb{E}\{|\textbf{v}_{i}^H\textbf{h}_k|^2\} - |\mathbb{E}\{\textbf{v}_k^H\textbf{h}_k\}|^2 +\sigma_n^2}.
\end{align}

In this study, we adopt coherent beamforming as the technique used to minimize the interference between the UTs. Therefore, the precoding coefficient for the $k$th UT and $m$th SAP is $v_{m,k}=\sqrt{p_{m,k}}\hat{h}_{m,k}$, where $p_{m,k}$ is a power scaling factor and $\hat{h}_{m,k}$ is the estimated UL channel, which is valid for the DL direction by virtue of channel reciprocity. That is, the precoding vector for the $k$th UT is given by
\begin{align}
\label{eq:precodingVector}
\textbf{v}_k=\textbf{P}_k^{1/2}\hat{\textbf{h}}_k,
\end{align}
where $\textbf{P}_k=\text{diag}\left(\frac{p_{1,k}}{\mathbb{E}\{|\hat{h}_{1,k}|^2\}},\frac{p_{2,k}}{\mathbb{E}\{|\hat{h}_{2,k}|^2\}},\cdots,\frac{p_{M,k}}{\mathbb{E}\{|\hat{h}_{M,k}|^2\}}\right)$ and $\hat{\textbf{h}}_k=[\hat{h}_{1,k},\hat{h}_{2,k},\cdots,\hat{h}_{M,k}]^T$ .

Accordingly, by using coherent beamforming as in (\ref{eq:precodingVector}),  phase-aware MMSE channel estimation as in (\ref{eq:estimatedChannel}), and the SINR in (\ref{eq:SINR}), the SINR can be derived as in  \cite{ozdogan2019performance} as follows:
\begin{align}
\label{eq:SINRCohBeam}
\text{SINR}_k^{\text{mmse}}=\frac{|\text{tr}(\textbf{P}_k^{1/2}\textbf{D}_k)|^2}{\text{Den} }, 
\end{align}
\begin{align}
\notag \text{Den} =& \sum_{k'=1}^K \text{tr}(\textbf{P}_{k'}\textbf{A}'_k\textbf{D}_{k'}) \\
\notag & + \sum_{k'\in\mathcal{C}_k\backslash k} q_kq_{k'}(\tau_u^p)^2|\text{tr}(\textbf{P}_{k'}^{1/2}\textbf{A}_k\textbf{G}_{k'}\textbf{A}_{k'})|^2 \\ &- \text{tr}(\textbf{P}_k\textbf{B}_k^2)+\sigma_n^2,
\end{align}
where
\begin{align}
\textbf{A}_k&=\text{diag}(\lambda_{1,k},\lambda_{2,k},\cdots,\lambda_{M,k}),\\
\textbf{A}'_k&=\text{diag}(\lambda'_{1,k},\lambda'_{2,k},\cdots,\lambda'_{M,k}),\\
\lambda'_{m,k} &= \lambda_{m,k}+\beta_{m,k},\\
\textbf{B}_k &= \text{diag}(\beta_{1,k},\beta_{2,k},\cdots,\beta_{M,k}),\\
\textbf{D}_k &= q_k\tau_u^p \textbf{A}_k\textbf{G}_k\textbf{A}_k + \textbf{B}_k,\\
\textbf{G}_k &= \text{diag}(\gamma_{1,k},\gamma_{2,k},\cdots,\gamma_{M,k})^{-1}.
\end{align}

Therefore, the achievable DL data rate (in bps/Hz) of the $k$th UT served by this cluster of SAPs can be calculated as 
\begin{align}
\label{eq:DLDataRate}
R_k=\frac{\tau_d^d}{\tau_c}\log_2\left(  1+ \text{SINR}_k \right)
\end{align}

\subsection{Cross-Layer Problem Formulation}
To optimize the power allocation and the handover decisions such that the cluster throughput is maximized and the handover rate is minimized, we formulate the power allocation and handover processes as a multi-objective optimization problem, where the objective functions to be maximized are the UTs' aggregate data rate and their service time before being switched to another cluster. For the latter, we maximize the number of served UTs with a guaranteed minimum data rate based on their link conditions. When the link condition does not allow optimized power allocation to serve the UT with the minimum required data rate, a handover request is issued. Then, a handover decision is taken when this repeats, and the visibility of the UT is confirmed by the next serving cluster, as discussed in Section \ref{sec:architecture}. That is, the two objectives of the optimization problem at the $t$th time slot are given as follows:
\begin{align}
\max_{p_{m,k},I_k}~~ \sum_{k=1}^K R_k[t]I_k[t]~~~\text{and}~~~\max_{p_{m,k},I_k}~~\sum_{k=1}^K I_k[t],
\end{align}
where $R_k[t]$ is the data rate of the $k$th UT during the $t$th time slot based on its channel conditions and power allocation during that time slot, as given in (\ref{eq:DLDataRate}). $I_k[t]$ is an indicator variable that indicates whether the $k$th UT can be served by the cluster during the $t$th time slot with an acceptable data rate, through optimizing the power allocation, or it is infeasible and a handover might be considered. If a handover is triggered, then the implementation procedure would be as shown in Fig.~\ref{fig:HOProcedure}. By introducing this indicator variable, we can jointly optimize the data rate of the UTs and their service time.

To deal with this multi-objective optimization problem, we construct a weighted sum of the two objectives to combine the two conflicting objectives into a single function. Thus, the complete handover-aware power allocation problem at time slot $t$ is formulated as follows:
\begin{align}
	\label{eq:OP1}
	&\max_{p_{m,k},I_k}~~ (1-\alpha) \sum_{k=1}^K R_k[t]I_k[t]+\alpha \sum_{k=1}^K I_k[t]\\	
	\tag{\theequation a}\label{eq:OP1a}\mathrm{s.t.}~~&R_k[t] \geq R_k^{\text{min}}I_k[t],~\forall k\in \mathcal{K}\\	
	\tag{\theequation c}\label{eq:OP1b} & \sum_{k=1}^K p_{m,k}\leq P_m^{\text{max}}, ~\forall m\in \mathcal{M}\\
	\tag{\theequation b}\label{eq:OP1c} &I_k[t] \in \{0,1\}, ~\forall k\in \mathcal{K}\\
	\tag{\theequation d}\label{eq:OP1d} & p_{m,k}\geq 0, ~\forall m\in \mathcal{M},~k\in\mathcal{K},
\end{align}
where $\alpha$ is a weighting coefficient that combines the two competing objectives and can be used to prioritize them. That is, by setting $\alpha=0$, we target maximizing the aggregate UTs' data rate only, and by setting $\alpha=1$, we target minimizing the handover rate only. Constraint (\ref{eq:OP1a}) is used to ensure that the served UTs satisfy their minimum rate level, where $R_k^{min}$ is the required minimum rate of UT $k$. Constraint (\ref{eq:OP1b}) is expressed to ensure that the total power scaling factors of every SAP are within the required range, where $P_m^{max}$  is the maximum total value of transmit power of the  $m$th SAP.  The binary value of the indicator variable, $I_k[t]$, and the non-negative value of $p_{u,k}$ are imposed by constraints (\ref{eq:OP1c}) and (\ref{eq:OP1d}), respectively.

Since the decision variables contain continuous (i.e., the power coefficient) and discrete (i.e., the handover indicator) ones and observing its structure, this optimization problem is modelled as a mixed-integer non-linear program (MINLP). This kind of optimization problem is known to be generally NP-hard due to its combinatorial behaviour \cite{Papadimitriou1998combinatorial}. Therefore, exponential computational complexity is required to solve the problem in (\ref{eq:OP1}). This means that solving this problem optimally cannot be done in real-time. In the next section, we leverage deep learning techniques to build a model that can predict efficient solutions to this problem in a computationally efficient manner. This is one of several benefits provided by AI techniques, as discussed in the next section.

\section{AI-Based Solution}
\label{sec:DLApproach}
In this section, we utilize data-driven techniques to implement the cross-layer control designed in Section \ref{sec:crosslayerDesign}. This can overcome the complexity of solving the optimization problem in (\ref{eq:OP1}) by moving most of the computations to be offline and brings the other benefits of AI to the satellite network management. For example, AI-based techniques adapt to the changing network conditions, which is one of the main characteristics of LEO SatNets. In addition, as multiple LEO satellites cooperate to serve the UTs as a cluster, distributed processing can be employed to reduce the processing load on the SSNs. In this regard, machine learning techniques support distributed implementation to a large extent \cite{hu2021distributed}. Therefore, implementing the proposed cross-layer control utilizing AI techniques can help in designing a more resilient satellite network.

Deep learning is one of the most efficient machine learning techniques, since it provides multi-layered models that are capable of learning efficient data representations from unstructured, complex datasets. Therefore, considering the optimized solution of the problem in (\ref{eq:OP1}) as the output of a certain mapping function, $f(\cdot)$, that maps the input of the optimization problem (i.e., user channel coefficients and QoS parameters), we design a deep neural network (DNN) model that can ``learn" this mapping function utilizing labelled training examples. For the latter, we use the traditional mathematical optimization tools to solve the problem in (\ref{eq:OP1}) offline for different inputs, and use it as the desired output of the DNN model. In doing so, we move the complexity of solving the optimization problem offline. We can then use the ``trained" DNN model to predict the optimized solution for new inputs without the need to mathematically solve the optimization problem again for the new inputs. It is worth noting that a trained DNN model can predict the output with much lower complexity compared to solving the optimization problem, since it is mainly executed as a  multiplication of matrices, as will be discussed in the sequel.  It should be also noted that model-free machine learning techniques (e.g., deep reinforcement learning) can be utilized to predict the solution of the problem in (\ref{eq:OP1}) by directly interacting with the network environment and learning from the decisions. However, by doing so, we would not utilize the labelled data that can be generated based on the slow-varying large-scale fading parameters. This is in addition to the other drawbacks of using model-free algorithms as discussed in \cite{arulkumaran2017deep}.
\begin{figure*}[t]
\centering     
\includegraphics[width=140mm]{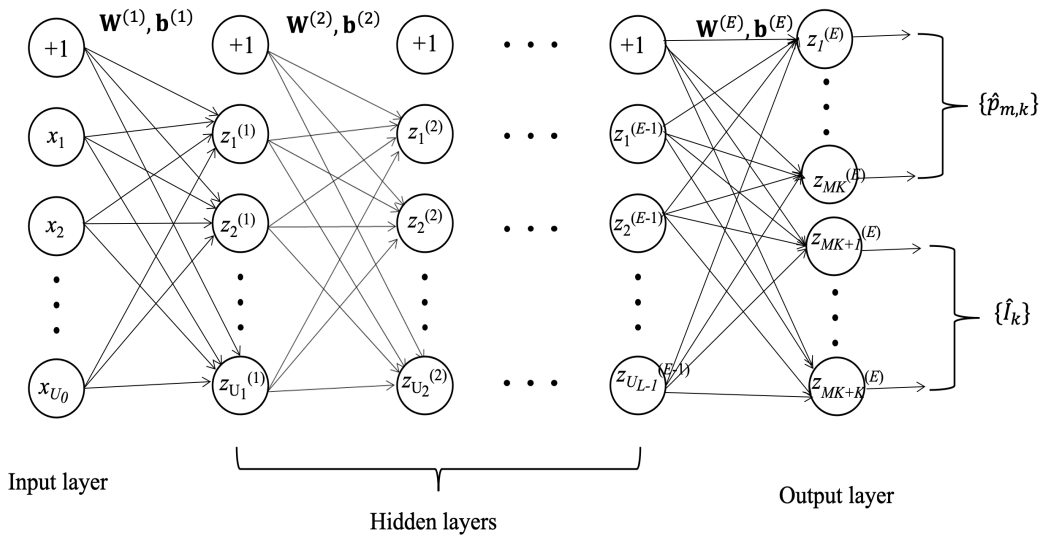}
\caption{The adopted DNN model. The input vector is based on (\ref{eq:DNNInputVect}) and the output is the prediction of the power allocation and handover indicator in (\ref{eq:desiredOutputVect}).}
\label{fig:DNNModel}
\end{figure*}

We determine the input and output vectors of the optimization problem in (\ref{eq:OP1}) as follows. The input vector is $\textbf{x}\in \mathbb{R}^{U_0},~U_0=MK+K+M$, and is given accordingly:
\begin{align}
\label{eq:DNNInputVect}
\textbf{x}=\left[\{L_{m,k}\}_{m=1,k=1}^{m=M,k=K},~\{R_k^{\text{min}}\}_{k=1}^{K},~\{P_m^{\text{max}}\}_{m=1}^{M}\right]^T.
\end{align}
This input is used by the DNN model to predict the desired output of the optimization problem, which is $\textbf{z}\in \mathbb{R}^{U_E},~U_E=MK+K$, and is calculated thus:
\begin{align}
\textbf{z} &= f(\textbf{x})\\
\label{eq:desiredOutputVect}&= \left[\{p^*_{m,k}\}_{m=1,k=1}^{m=M,k=K},~\{I_k^*\}_{k=1}^{K}\right]^T,
\end{align}
where $f(\cdot)$ is the function that represents the mapping process of the optimization problem in (\ref{eq:OP1}), and $p^*_{m,k}$ and $I_k^*$ are the desired values of power allocation and handover indicator, respectively. Therefore, the output of the DNN model is the predicted value for this desired vector, $\textbf{z}$. That is, $\hat{\textbf{z}} = \hat{f}(\textbf{x})$, where $\hat{f}(\cdot)$ is the approximated mapping function.

The adopted DNN model is shown in Fig. \ref{fig:DNNModel}. The DNN model is composed of $E$ layers that include an output layer and $E-1$ hidden layers. Each layer has a number of neurons or units. Assume that the $e$th layer incorporates $U_e$ units. The output of the DNN model, $\hat{\textbf{z}}$, can be calculated using a multi-layered matrix multiplication process. That is, the output of the $e$th layer can be calculated as follows:
\begin{align}
\label{eq:outputPerLayer}
\textbf{z}^{(e)} = g^{(e)}\left( \textbf{W}^{(e)} \textbf{z}^{(e-1)} + \textbf{b}^{(e)} \right),
\end{align}
where $\textbf{z}^{(e)}$ is the output of the $e$th layer (i.e., $\hat{\textbf{z}} = \textbf{z}^{(E)}$) and $g^{(e)}(\cdot)$ is the activation function that maps the output of the previous layer to the input of the subsequent one. In addition to the activation function, the parameters $\textbf{b}^{(e)}$ and $\textbf{W}^{(e)}$ are used to calculate the input of the activation function based on the output of the previous layer, where $\textbf{b}^{(e)}\in \mathbb{R}^{U_e}$ is the bias vector of the $e$th layer and $\textbf{W}^{(e)}\in \mathbb{R}^{U_e\times U_{e-1}}$ is its weights matrix such that $w^{(e)}_{ij}$ is the weight of the edge from the $j$th unit in layer $e-1$ to the $i$th unit in layer $e$. These parameters ($\textbf{b}^{(e)}$ and $\textbf{W}^{(e)}$) are optimized to provide accurate predictions for the desired output vector. This optimization process is implemented during the training phase.

To train the DNN model (i.e., to optimize $\textbf{b}^{(e)}$ and $\textbf{W}^{(e)}$), a labelled dataset is constructed by using the optimal solution of the optimization problem in (\ref{eq:OP1}) at different inputs. It is worth mentioning that, alternatively, a suboptimal solution can be used, in case the optimal solution is too expensive to calculate. However, the DNN model would mimic the suboptimal procedure. Assume that the training set is composed of $T$ tuples $\{(\textbf{x}^{\{1\}},\textbf{z}^{\{1\}}),(\textbf{x}^{\{2\}},\textbf{z}^{\{2\}}),(\textbf{x}^{\{3\}},\textbf{z}^{\{3\}}),\cdots,\newline (\textbf{x}^{\{T)},\textbf{z}^{\{T\}})\}$. Therefore, the DNN model is trained by solving the following unconstrained optimization problem:
\begin{align}
\min_{\{\textbf{W}^{(e)}\},\{\textbf{b}^{(e)}\}}~~ \frac{1}{T} \sum_{i=1}^T \mathcal{L}\left( \textbf{z}^{\{i\}},\hat{\textbf{z}}^{\{i\}} \right),
\end{align}
where $\mathcal{L}(\cdot)$ is a loss function that is selected to represent the error between the prediction and the ground truth of the training examples. 

After the offline training phase, the DNN model becomes ready to provide efficient predictions for the optimized power allocation and handover indicator online based on the new input. The new input is mainly composed of the values  of the large-scale parameters at each time instant. This online operation does not need re-training the DNN model for those new values as the DNN model learns from the training examples during the training phase.

\subsection{Complexity Analysis}
\label{ssec:complexityAnalysis}
During online operation, the DNN model calculates predictions for the desired output vector for the input values based on the experience gained in the training phase. For this purpose, the output of each layer is calculated as in (\ref{eq:outputPerLayer}). Therefore, the computational complexity to calculate $\textbf{z}^{(e)}$ is due to the multiplication operation and the activation function. Using the big-O notation, this computational complexity is in the order of $\mathcal{O}(U_e U_{e-1 }+ U_e)=\mathcal{O}(U_e U_{e-1 })$. Therefore, the complexity of calculating the output of the DNN model is  $\sum_{e=1}^E \mathcal{O}(U_e U_{e-1 })$. For example, if two hidden layers are used, and each layer has the same number of units as the input, then the computational complexity can be calculated as follows:
\begin{align}
\sum_{e=1}^3 \mathcal{O}(U_e U_{e-1 }) &= \mathcal{O}(U_0^2) + \mathcal{O}(U_0^2) + \mathcal{O}(U_0U_E)\\
&= \mathcal{O}(U_0^2) + \mathcal{O}(U_0U_E)\\
&= \mathcal{O}((MK+K+M)^2) + \\
\notag &\mathcal{O}((MK+K+M)(MK+K))\\
&= \mathcal{O}(M^2K^2).
\end{align}
This is much lower than the exponential complexity of solving an NP-hard problem, such as (\ref{eq:OP1}). In addition, other benefits of using machine learning-based techniques are utilized, such as adaptability to the dynamics of the satellite link and topology, scalability, and support of distributed data processing and storage.

%
\section{Simulation Results}
\label{sec:Results}
\begin{table}[t]
\renewcommand{\arraystretch}{1.3}
\caption{Simulation Parameters}
\label{tab:simParameters} 
\centering
\begin{tabular}{llc}
\hline 
\textbf{Parameter} & \textbf{Value} \\ 
\hline 
Satellites altitude & $550$ km \\
Antenna factor ($\eta$) & 20 \cite{li2020hierarchical}\\
Carrier frequency & $30$ GHz \\
Shadowing std & $5$ dB \\
Noise figure & $7$ dB \\
Noise power spectral density & $-174$ dBm/Hz \\
Sat. max power ($P_m^{\text{max}}$) & $15$ dBW\\
Sat. antenna gain & $30$ dB\\
UT antenna gain & $5$ dB \\
Pilot power ($q_k$) & $5$ dBW \\
Coherence intervals: $\tau_c,~\tau_u^p$ & $300,~30$ samples\\
Number of runs & $10$ \\
Number of UTs ($K$) & $10$ \\
Distribution of UTs & Uniform\\
Number of antennas per satellite  & 1 \\
Priority factor ($\alpha$) & $0.5$ \\
\hline
\end{tabular}
\end{table}
Here, we present and discuss the simulation results to evaluate the performance of the proposed DM-MIMO-based architecture, optimized cross-layer design, and AI-based implementation. In addition, we compare the performance with that of baseline techniques from the literature. 

We consider a set of $M$-LEO satellites that serves a set of UTs distributed uniformly over a  $1,000\times 1,000~\text{km}^2$ area. The values of the adopted simulation parameters are summarized in Table \ref{tab:simParameters}. For the proposed D-JPAHM approach, this set of satellites is considered a cluster of SAPs, and each UT is connected to the whole cluster in a DM-MIMO manner. The CPU directs the UTs' downlink data to the SAPs and determines the power allocation for each SAP-UT link according to the optimization problem in (\ref{eq:OP1}). Without loss of generality, we assume that conjugate beamforming is used to determine the beamforming vectors, as been used in several studies in the literature (e.g.,\cite{ngo2017cell,chen2018channel,ozdogan2019performance}). To solve this optimization problem, we use Matlab's genetic algorithm (GA) solver. It should be noted that the sub-objectives of the problem in (\ref{eq:OP1}) should be normalized. This normalization along with the weighting factor ($\alpha$) ensure that the combined objectives are on the same scale. As a baseline, we compare the proposed DM-MIMO-based architecture with  traditional single-satellite connectivity. For the handover scheme, we consider two well-known techniques. In the first scheme, each UT connects to the LEO satellite with the best channel condition and switches to a different one when the new satellite has a better link condition \cite{wu2016graph}. This approach maximizes the throughput of each UT and is referred to as \texttt{BestChannel}. In the second single-satellite connectivity technique, each UT remains connected to the LEO satellite in its visibility as long as its achievable rate is higher than  the minimum acceptable value $R_k^{\text{min}}$. When this minimum value can not be achieved by the current satellite connection, the UT switches to the best channel LEO satellite \cite{wu2016graph}. This technique maximizes the service time for each UT (i.e., minimizes its handover rate) and is referred to as \texttt{MaxServTime}.

\begin{figure}[t!]
 \centering
 \includegraphics[width=0.99\linewidth]{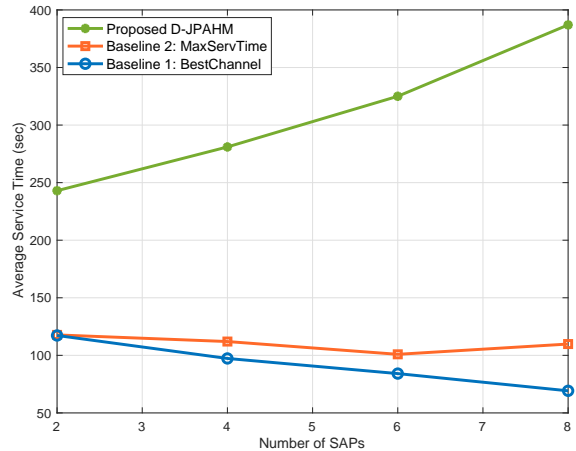}
 \caption{Average service time of the UTs versus the number of SAPs. The proposed D-JPAHM technique achieves the highest average service time compared to single-satellite-based approaches.}\label{fig:AvrgeServTime}
 \end{figure}
 \begin{figure}[t!]
 \centering
 \includegraphics[width=0.99\linewidth]{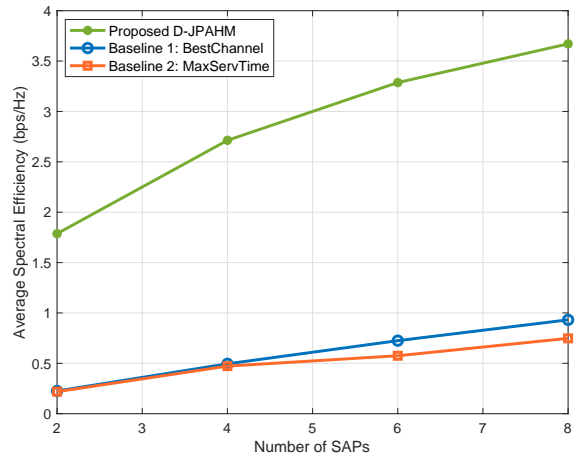}
 \caption{Average spectral efficiency of the UTs versus the number of SAPs. The proposed D-JPAHM scheme outperforms traditional techniques.}
\label{fig:AvrgeSE}
\end{figure}

The average service time of the UTs is plotted against the number of satellites, $M$, using the three techniques in Fig.~ \ref{fig:AvrgeServTime}. As we can see, the \texttt{MaxServTime}  technique achieves higher service time than  \texttt{BestChannel}, which is as we would expect since \texttt{MaxServTime} prioritizes the longer service time, not the achievable data rate. However, the proposed D-JPAHM approach achieves a much higher value given that the visibility of a cluster of satellites is always longer than that of a single satellite. Therefore, the average service time increases with the increase of the number of SAPs in the cluster.

Fig.~\ref{fig:AvrgeSE} shows the average spectral efficiency (in bps/Hz) versus the number of SAPs using the three approaches. As we can see, the cooperative transmission, full frequency reuse, and optimized power allocation of the DM-MIMO-based architecture improved spectral efficiency and outperformed conventional single-satellite connectivity techniques. Furthermore, this spectral efficiency gain increased with the increase of the number of SAPs. Although \texttt{BestChannel} prioritized spectral efficiency, it could not compete with the DM-MIMO architecture, as shown in the figure. This spectral efficiency improvement could be exploited towards direct broadband connectivity of handheld devices.

\begin{table}[t]
\renewcommand{\arraystretch}{1.3}
\caption{DNN Model Parameters}
\label{tab:DNNParameters} 
\centering
\begin{tabular}{llc}
\hline 
\textbf{Parameter} & \textbf{Value} \\ 
\hline 
Dataset size & $33,000$ samples \\
Training dataset percentage & $70\%$\\
Optimizer & Adam (learning rate = $0.01$)\\
Batch size & $32$ \\
Number of hidden layers & $2$ \\
Number of units per layer & $40,~50$ \\
Activation functions & Hard-sigmoid\\
Loss functions & MSE, Binary Cross-entropy\\
Training epochs & $50$ \\
\hline
\end{tabular}
\end{table}

To evaluate the accuracy of the proposed DNN model discussed in Section \ref{sec:DLApproach}, we generated a dataset with simulations that are implemented in Matlab. We simulate a cluster of four satellites that serves a set of six UTs, which were uniformly distributed in the service area, and the positions were changed randomly at each simulation run  to generate a total of $33,000$ samples of channel realizations for the UTs. The simulation parameters are summarized in Table \ref{tab:simParameters}. Given the channel conditions of the UTs at every time instant, the optimization problem in (\ref{eq:OP1}) was solved using Matlab's GA solver to construct the input and corresponding output of the dataset, as discussed in Section \ref{sec:DLApproach}. This offline-generated dataset was then used for the training and testing of the DNN model. The DNN model was implemented and tested on Python using TensorFlow and Sklearn packages. The DNN model parameters are summarized in Table \ref{tab:DNNParameters}.

To evaluate the accuracy of the predictions of the DNN model, we used the mean squared error (MSE) of the predicted power allocation (normalized to $P_m^{\text{max}}$) and the F1 score for the prediction of the handover indicator. The F1 score is an efficient metric that is used widely in data science and binary classification problems \cite{huang2015maximum}. The F1 score can be calculated using the following expression:
\begin{align}
\text{F1 Score} = \frac{2\times \text{Precision}\times \text{Recall}}{\text{Precision}+ \text{Recall}},
\end{align}
where
\begin{align}
\text{Precision} = \frac{\text{True Positives}}{\text{True Positives}+ \text{False Positives}},\\
\text{Recall} = \frac{\text{True Positives}}{\text{True Positives}+ \text{False Negatives}}.
\end{align}

\begin{figure}[t!]
 \centering
 \includegraphics[width=0.99\linewidth]{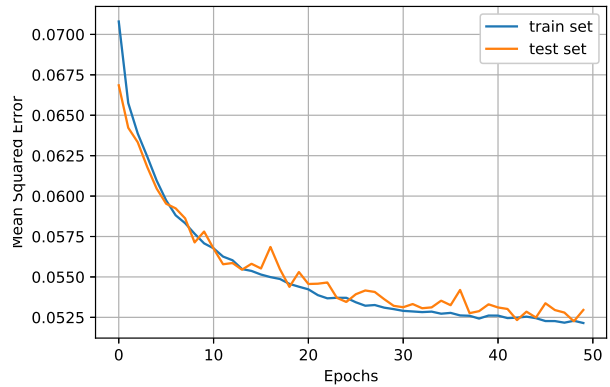}
 \caption{Mean squared error (MSE) of the power allocation prediction for the training and testing datasets against the number of epochs. The MSE of the testing dataset approaches that of the training dataset, and both achieve high prediction accuracy.}\label{fig:Power_mse_learncurve}
\end{figure}

Figure \ref{fig:Power_mse_learncurve} shows the MSE of the predicted power allocation $\{\hat{p}_{m,k}\}$ (normalized to $P_m^{\text{max}}$) versus the number of training epochs to represent the learning curve of the DNN model. This was plotted for both the training and testing datasets. As we can see, the predictions achieved a high accuracy compared to the optimized solutions at the end of the training process. In addition, the gap between the testing and training MSE was small, which means that there is no overfitting problem in this DNN model. 

\begin{figure*}[t!]
\centering     
\subfloat[]{\label{fig:I_precision_learncurve}\includegraphics[width=3.2in]{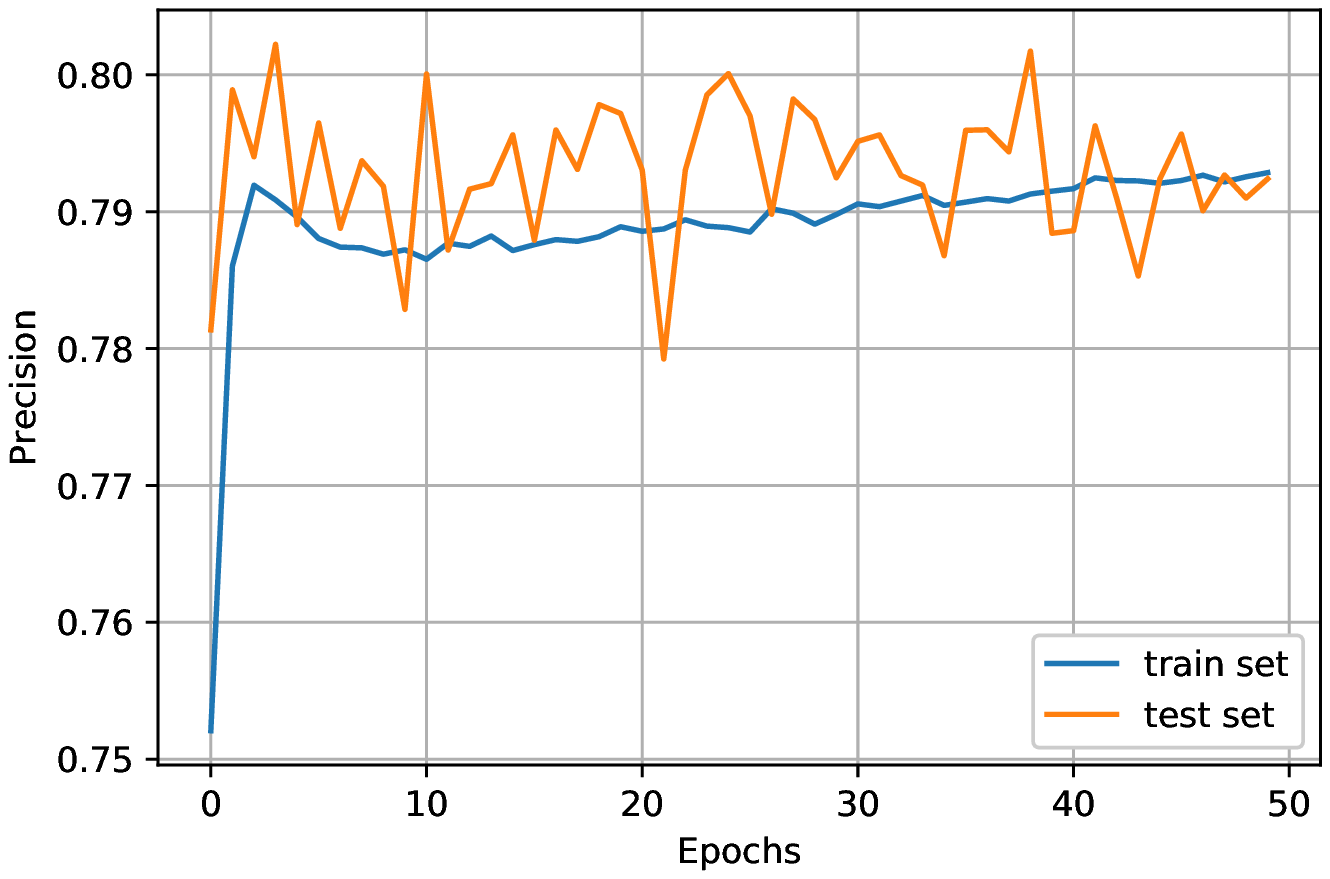}}
\hfil\subfloat[]{\label{fig:I_recall_learncurve}\includegraphics[width=3.2in]{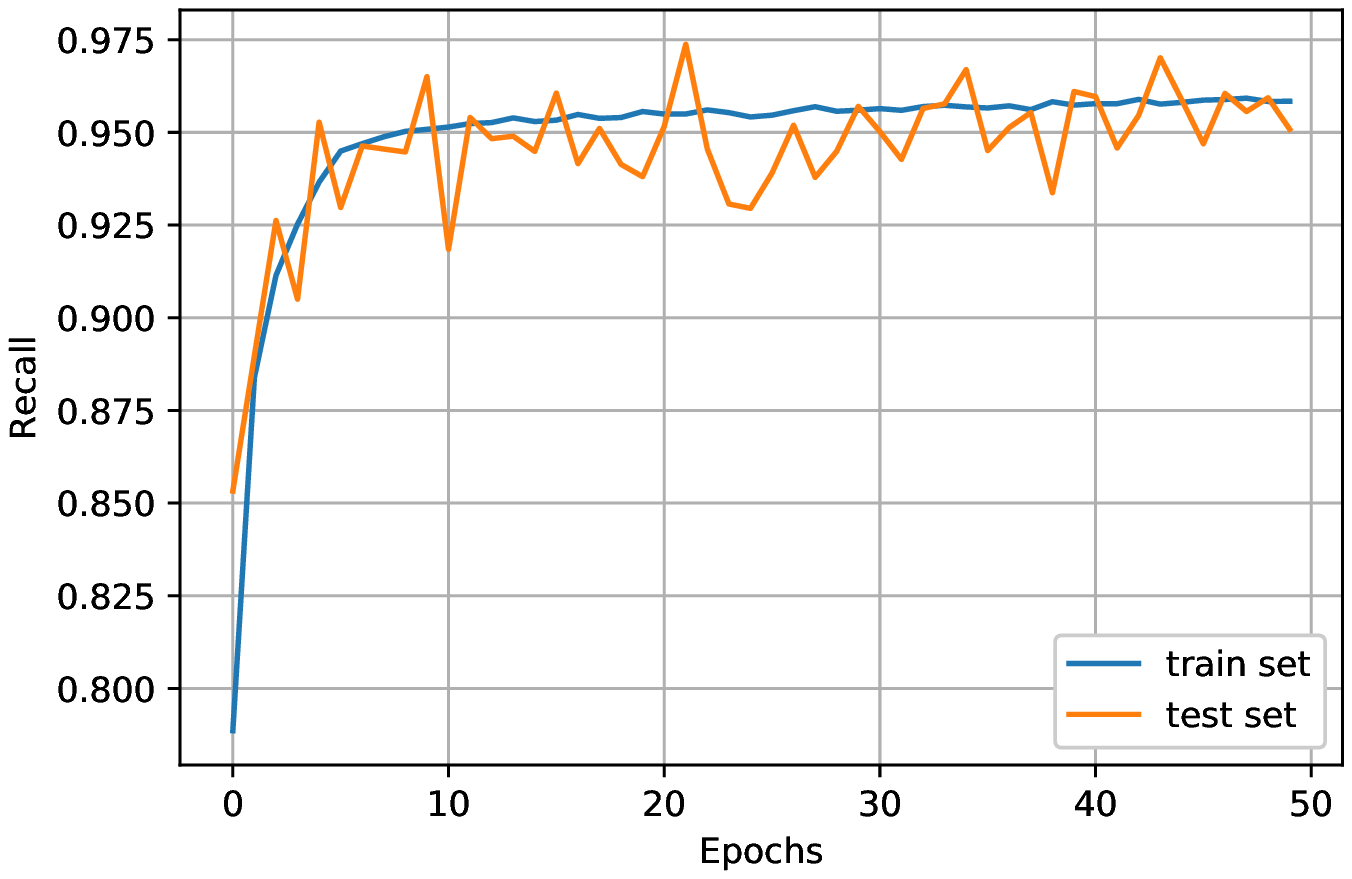}}
\hfil\subfloat[]{\label{fig:I_f1_learncurve}\includegraphics[width=3.2in]{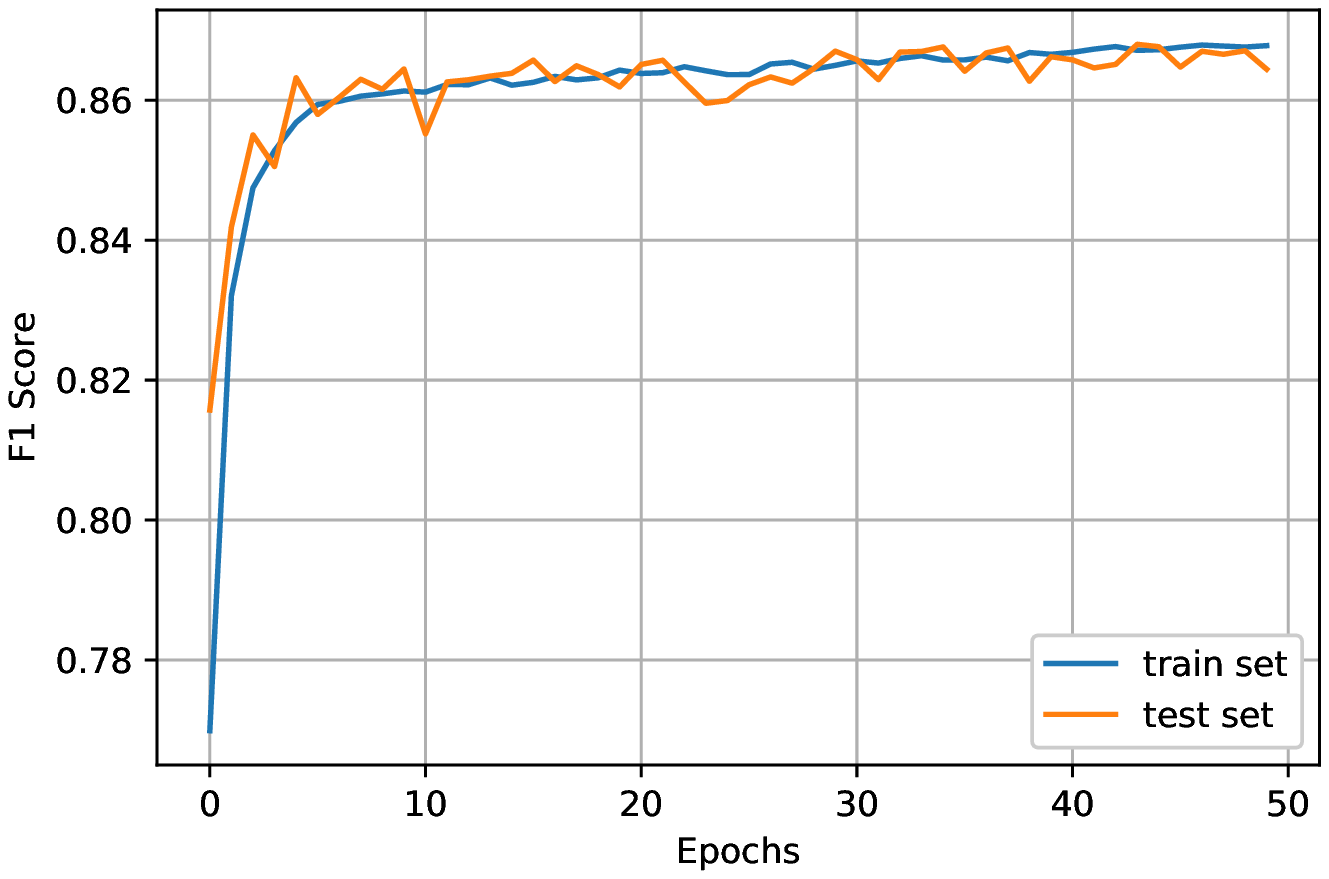}}
\caption{The accuracy of the handover indicator prediction in terms of Precision in (a), Recall in (b), and the overall F1 Score in (c) for the training and testing datasets against the number of epochs. The DNN model provides predictions for the handover indicator with high accuracy. }
\label{fig:I_predictionAccuracy}
\end{figure*}

Similarly, the Precision, Recall, and F1 score of the predicted handover indicator, $\{\hat{I}_{k}\}$, were plotted against the training epochs in Fig.~\ref{fig:I_predictionAccuracy} to evaluate the accuracy of the prediction. Again, the DNN model achieved accurate predictions for the handover process. Therefore, this shows that the DNN model can effectively mimic the implicit mapping function of the optimized cross-layer design developed in Section \ref{sec:crosslayerDesign}. However, it operates online with much lower complexity compared to that of  solving the NP-hard problem in (\ref{eq:OP1}) using traditional mathematical optimization techniques. This means that the proposed cell-free, optimized cross-layer design can be executed in practice utilizing deep learning techniques.
%

%
\section{Conclusions and Future Work}
\label{sec:Conclusions}
In this paper, we introduced a novel DM-MIMO-based LEO satellite network architecture and discussed its various design aspects, benefits, associated challenges, and potential solutions. Based on the proposed architecture, we developed an optimized cross-layer control framework, where the power allocation and handover management processes were jointly optimized. In addition, we introduced an AI-based implementation suitable for real-time operation and the dynamic environment of the LEO satellite network. The simulation results demonstrated that the proposed DM-MIMO-based architecture achieves better spectral efficiency for the network than baseline techniques and reduces the handover rate of the user terminals, exploiting the ultra-dense deployment of the LEO satellites. Moreover, the results showed that the deep learning-based implementation can predict efficient and accurate solutions for the proposed optimization framework. Therefore, the proposed architecture and solutions can enable future ultra-dense LEO satellite networks to be more efficient, resilient, and intelligent. 

In future work, we will investigate the other aspects of system design based on this architecture, such as optimizing the clustering process, the synchronization between transmissions, and the outdated channel information. The potential solutions that were proposed in the literature for these challenges were discussed in Section \ref{ssec:ArchChallenges}. However, further optimization needs to be investigated.

%
\ifCLASSOPTIONcompsoc
  \section*{Acknowledgments}
\else
  \section*{Acknowledgment}
\fi
This work has been  supported by the National Research Council Canada’s (NRC) High  Throughput Secure Networks program (CSTIP Grant \#CH-HTSN-607) within  the Optical Satellite Communications Consortium Canada (OSC) framework.
%

%
%
\bibliographystyle{ieeetr}
\bibliography{References}
\end{document}